\documentclass[showkeys,showpacs,amsmath,superscriptaddress,preprintnumbers,amssymb,10pt,twocolumn]{revtex4-2}
\usepackage{graphicx}
\usepackage{bm}
\usepackage{color}
\usepackage{subfigure}
\usepackage{amssymb}
\usepackage{ulem}
\begin{document}
\title{Optical solitons in  saturable cubic-quintic nonlinear media with nonlinear dispersion}
\author{Sudipta Das}
\email{das2179@gmail.com}
\affiliation{Department of Physics, Govt General Degree College, Chapra, WB, India}
\author{Kajal Krishna Dey}
\affiliation{Department of Physics, B. B. College, Asansol, WB, India}
\author{Golam Ali Sekh}
\email{skgolamali@gmail.com}
\affiliation{Department of Physics, Kazi Nazrul University, Asansol-713340, India}
\begin{abstract}
We study  an intense-short pulse propagation in a saturable cubic-quintic nonlinear media in the presence of  nonlinear dispersion within the framework of an extended variational approach. We derive an effective  equation  for the pulse width and demonstrate how the saturation due to nonlinearity is achieved in presence of nonlinear dispersion.  We find that the nonlinear dispersion can change the pulse width and induce motion in the system. The direction of induced motion depends on the sign of nonlinear dispersion. The pulse is energetically stable at an equilibrium width. A disturbance can, however, induce oscillation in pulse width, the frequency of which is always smaller due to nonlinear dispersion. We check dynamical stability  by a direct  numerical  simulation. 
\end{abstract}
\maketitle
\section{Introduction}
Studies on the propagation of optical pulse in nonlinear  media have been receiving a great attention in the context of solitary-wave-based communication, especially, in femtosecond domain and, consequently, in ultrafast optics (pulse compression) and all-optical switching for last few decades. While propagating several nonlinear phenomena including filamentation, harmonic generation, third-order dispersion, self-steepening  and self-frequency shift are associated with the pulse. The observations of such nonlinear phenomena depend crucially on the properties of the medium and the pulse \cite{rrr1,rrr2}.

An intense  short pulse  induces  higher-order nonlinearities (HON) in an optical medium.  Effects of such pulse are described  by various forms of  generalized nonlinear Schr\"odinger equation(GNLSE). The GNLSE  is a version  of the NLS equation which  includes the different types nonlinear terms, some of which  have already been realized experimentally. For example, quintic nonlinearity has been obtained  in semiconductor doped glasses while  septic  nonlinearity have been measured in different glasses. It can also be possible to obtain saturation in the nonlinearity in some materials if  the intensity of the pulse is relatively high. In semiconductor-doped glasses and organic polymers, however, the nonlinearity becomes saturated even at moderate pulse intensities \cite{rr00,rr001,rr002}.

Self-steepening  is one of the effects that  commonly  arises due to propagation of ultrashort intense pulse in a nonlinear medium. The self-steepening  is related to  nonlinear dispersion  \cite{r001}. In the presence of this effect several attempts have been made to analyse the response  of higher-order nonlinearities on the propagation of such ultrashort pulse\cite{rr1,r01}.  In the recent past,  the existence and stability of various types of  wave patterns (periodic pattern, bright  and dark solitary pulses) have been investigated by  Chow et al. It is seen that these modes are stablized due to competition  among different higher-order nonlinearities \cite{rr2}.  The existance of chirped solitary wave solution  has been investigated by Triki et al\cite{rr3}.  Recently,  Konar et al studied the characteristics of chirped solitary pulse in dispersive media with cubic saturable nonlinearity and showed that the pulse broadening can be reduced with increase of saturation limit \cite{rr4}. The effect of cubic-quintic saturable nonlinearity on the fundamental bright soliton have been studied in dispersive media in Ref.\cite{rr5,rr7}.

{Our objective in this work is to investigate the properties of bright  optical soliton  in the presence of nonlinear dispersion and saturable cubic-quintic nonlinearity, specifically,  (i) how does a parameter of soliton  reaches its saturable limit? and (ii) what are the changes  in  the static and dynamical properties of soliton induced by nonlinear dispersion?} In the  zero nonlinear dispersion (ND) limit,  the system  supports stable localized solitary waves due to saturable cubic-quintic nonlinearity. These solitons are dynamically stable.  The effect of non-zero ND on the pulse  is  described by a derivative nonlinear Schr\"odinger equation (DNSE).  We work with the variational approach and find an effective potential for the pulse width in the parameter space. Treating the problem in terms of effective potential in the parameter space is often termed as a potential model. This model are widely used to predict  static and dynamical properties of the solitary waves in the context of nonlinear optics \cite{r1,rr1a,rr0008,rr0009,rr0010} and  Bose-Einstein condensates \cite{r11,r11a,r111}.  

In section II, we introduce the mathematical model and numerically predict the existence of localized bright solution in the limit of  negligible nonlinear dispersion. With a view to find the effect of nonlinear dispersion on the solitary wave we work with Ritz optimization procedure in section III. We derive  equations for the different parameters of the solitonic pulse  and analyse the effect of saturable nonlinearity on it. In section IV we present the result on the properties of propagating pulse considering nonlinear dispersive effect. We devote section V to make some concluding remarks. 
\section{Theoretical Model}
We consider  propagation of an intense  pulse in optical waveguide and model the evolution of pulse envelope ($\psi(z,\tau)$)  by the derivative non-linear Schr\"odinger equation \cite{rr2,rr3,rr4,rr5,rr6}
\begin{eqnarray}
	i\psi_z&=&\alpha\frac{d^2\psi}{d\tau^2}+i\gamma|\psi|^2\frac{d\psi}{d\tau}+\kappa|\psi|^2\psi+\frac{\beta\,|\psi|^4\psi}{(1+G|\psi|^4)}\nonumber\\
	&+&\frac{\mu \,|\psi|^2\psi}{(1+ S\,|\psi|^2)}
\label{eq1}
\end{eqnarray}
Here $\alpha$, $\gamma$, $\kappa$, $\beta$ and $\mu$ stand for strength of  group velocity dispersion, self-steepening, cubic nonlinearity, saturated quintic and cubic nonilinearities respectively. For $ 0< G \le 1$ the 4th term in Eq. (\ref{eq1}) can be expanded as: $|\psi|^4\psi-G|\psi|^8\psi+\cdots$. It is called saturable quintic nonlinear term. Similarly, the term for saturated cubic nonlinearity  can be expanded for $ 0< S \le 1$.   In this model we have considered the dispersive effect upto second-order. This type of dispersion management is feasible in the manufacturing industry by adjusting different parameter/ingredient of an optical fiber.

\begin{figure}[h]
\centering
\includegraphics[width=0.34\textwidth]{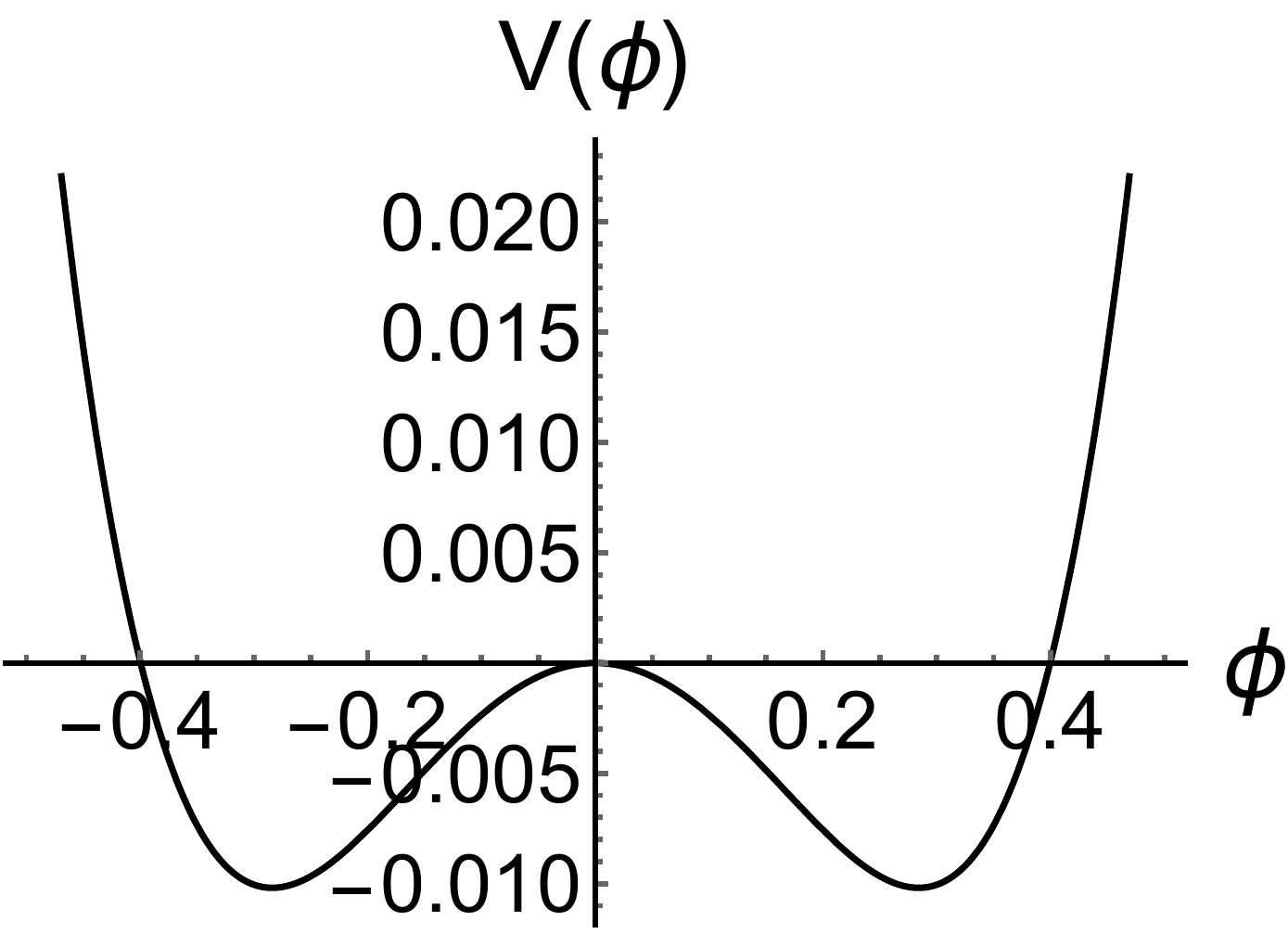}
\includegraphics[width=0.35\textwidth]{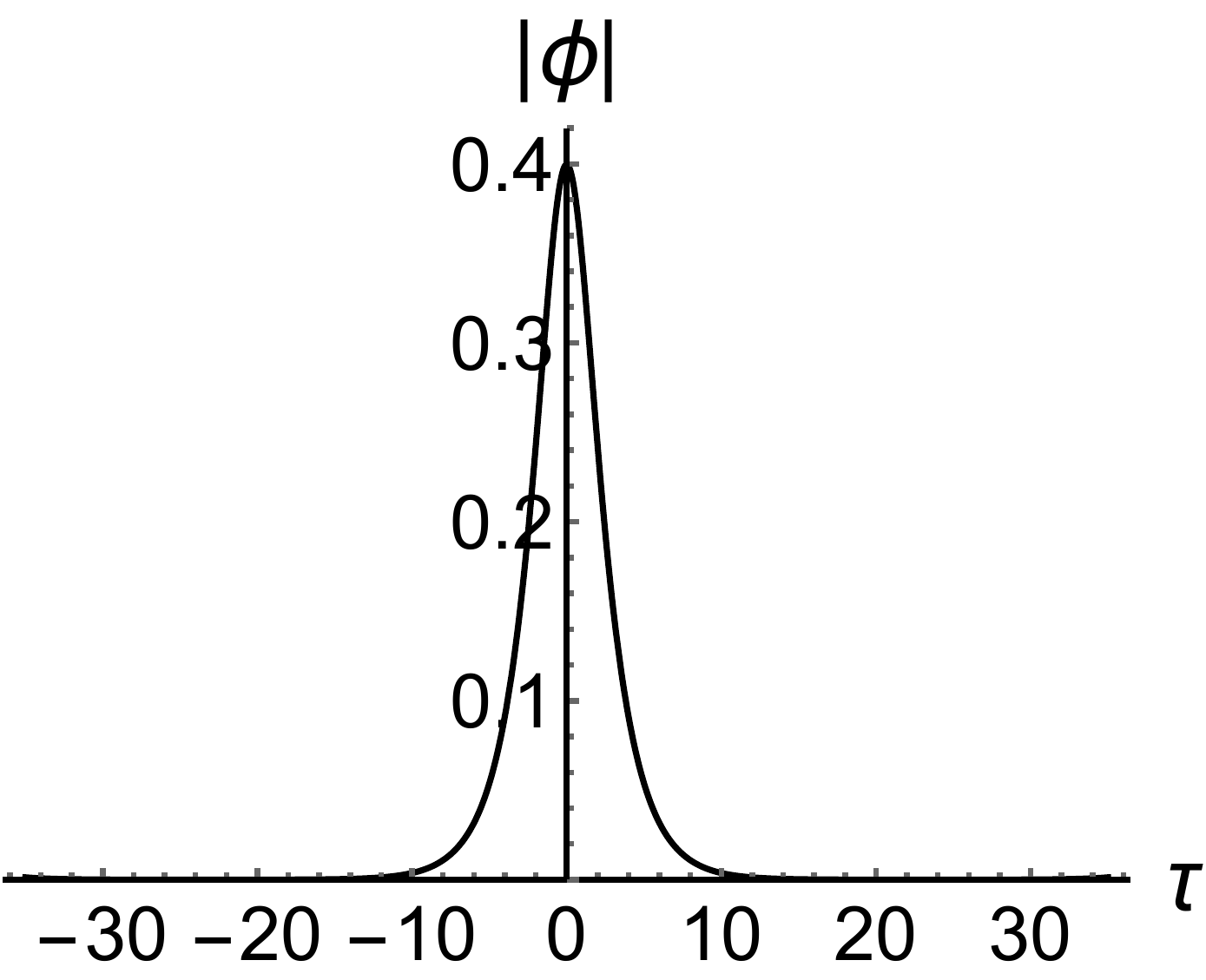}
\caption{Top panel: The potential $V(\phi)$ as a function of $\phi$ for $\kappa=4$, $\mu=2$, 
$\beta=3$ and $\omega=0.5$ for bright type profile in a saturable quintic and cubic nonlinear media with $G=0.27$ and $S=0.27$. Bottom panel:  Stationary solution of NLS with saturable quintic nonlinearity for $\alpha=1.7$ with the initial condition $\phi(0)=0.399605$ and $\phi'(0)=0$. }
\label{Fig1}
\end{figure}

In order to understand the properties  of stationary solution of Eq. (\ref{eq1}) in the $\gamma \rightarrow 0$ limit,  we introduce $\psi(z,\tau)=\phi(\tau)\,\exp(-i \omega z)$ and write z-independent GNLS equation with saturable nonlinearities 
\begin{eqnarray}
\omega {\phi}=\alpha\frac{d^2{\phi}}{d{\tau}^2}+\kappa|\phi|^2\phi+\frac{\beta\,|\phi|^4\phi}{\left(1+ G|\phi|^4\right)}+\frac{\mu \,|\psi|^2\psi}{\left(1+ S|\psi|^2\right)},\,\,\,
\label{eq2}
\end{eqnarray}
the  energy functional for which is given by
\begin{eqnarray}
E[\phi]=\frac{\alpha}{2}\left( \frac{d\,\phi}{d\tau}\right)^2+V(\phi)
\label{eq3}
\end{eqnarray}
with
\begin{eqnarray}
V(\phi)&=&-\frac{\beta {\rm tan}^{-1}\left(\sqrt{G} \phi^2\right)}{2 G^{3/2}}-\frac{(-\beta +G\, \omega ) \phi^2}{2 G}\nonumber\\&+&\frac{1}{4} \kappa \phi^4-\frac{\mu  \ln \left(1+S \phi^2\right)}{2 S^2}+\frac{\mu  \phi^2}{2 S}
\label{eq4}
\end{eqnarray}
Here $V(\phi)$ is a double-well shaped potential.  It has one positive stationary point and one zero stationary point (top panel Fig. \ref{Fig1}). The stationary point at $\phi=0$ corresponds  to the fundamental  of solitary-wave solution: $\phi(\tau)\,\rightarrow\, 0$ as $\tau\,\rightarrow\,\infty$.  Considering $\phi'(0)=0$ if we take $\phi(0)\ge \phi$ for solving Eq.(\ref{eq2}), we can obtain different modes of solutions\cite{rr8}. With a view to confine our attention to fundamental mode of solution we take $\phi=\phi(0)+\Delta \phi$ and solve of Eq. (\ref{eq2}) numerically. The  result is displayed in the bottom panel of Fig. \ref{Fig1}. We see that the NLS  with saturable cubic-quintic nonlinearity (SCQNL) can support bright type solitary profile. 
It is relevant to note that for $\gamma \neq 0$, the existence of chirped solitary profile in the presence of unsaturated nonlinearity  has recently been analysed  in Ref.\cite{rr2,rr3}.
\section{Variational formulation}
We have seen that the stationary solution of Eq. (\ref{eq1}) supports a single hump bright soliton. In order to understand the effect of nonlinear dispersion on the dynamics of optical pulse propagating in a quintic saturable nonlinear medium we consider variational approach.  To begin with we follow inverse variational method and write Lagrangian density ${\cal L}$ for Eq. (\ref{eq1})  as
\begin{eqnarray}
{\cal L}&=&i(\psi\psi_z^*-\psi^*\psi_z)+\frac{1}{2}i\gamma|\psi|^2(\psi_\tau\psi^*-\psi_\tau^*\psi)-2\alpha|\frac{d\psi}{d\tau}|^2\nonumber\\&+&\kappa|\psi|^4+\frac{2\beta}{G}|\psi|^2-\frac{2\beta \,{\rm tan}^{-1}(\sqrt{G}|\psi|^2)}{G^{3/2}} +\frac{2\mu}{S}|\psi|^2\nonumber \\
&-&\frac{2\mu}{S^2}\ln(1+ S \,|\psi|^2)
\label{eq5}
\end{eqnarray}
and we consider 
\begin{eqnarray}
\psi(z,\tau)&=&B \,{\rm sech}\left[\frac{\tau-\tau_0}{w}\right] \times \nonumber \\  &&\exp{i\left[v(\tau-\tau_0)+\frac{q}{2w}(\tau-\tau_0)^2+\sigma\right]}\,\,\,\,\,
\label{eq6}
\end{eqnarray}
as a trial solution of Eq.(\ref{eq1}) for the bright density profile with the norm $P=\int |\psi(z,\tau)|^2\,d\tau=2 w |B|^2$. Here, $B$, $\tau_0$ and $w$ represent respectively the complex amplitude, central position, pulse width of the $\psi$ envelope. Other parameters, namely, $\sigma$, $v$ and $q$ indicate the phase, velocity (center of the soliton) and frequency chirp respectively. In the variational analysis, we allow all these parameters to vary with $z$  with view to catch effects of the system \cite{rr0007}.

Inserting Eq.(\ref{eq6}) in Eq.(\ref{eq5}) and then integrating the resulting equation with respect to $\tau$ from $-\infty$ to $+\infty$ we obtain  the following averaged Lagrangian density 
\begin{eqnarray}
\langle {\cal L}\rangle&=&2i\left(B\frac{dB^*}{dz}-B^*\frac{dB}{dz}\right)w+2|B|^2\left(-2vw\frac{d\tau_0}{dz}\right.\nonumber\\ &+&\left.\frac{\pi^2}{12}w^2\frac{dq}{dz}-\frac{\pi^2}{12}qw\frac{dw}{dz}+2w\frac{d\sigma}{dz}\right)-\frac{4}{3}\gamma|B|^4wv\nonumber\\&-&4\alpha|B|^2\left(\frac{1}{3w}+v^2w+\frac{\pi^2}{12}q^2w\right)+\frac{4}{3}\kappa|B|^4w\nonumber\\&+&\frac{4\beta}{G}|B|^2w
-\beta M|B|^{2(2n-1)}w+\frac{4\mu}{S}|B|^2w\nonumber\\&-&\mu N|B|^{2n}w,
\label{eq7}
\end{eqnarray}
where
\begin{eqnarray*}
M&=&\frac{2}{G^{3/2}}\sum_{n=1}^{\infty}\frac{(-1)^{n+1}}{2n-1}G^{\frac{2n-1}{2}}\left(M_1+M_2\right)
\end{eqnarray*}
with
\begin{eqnarray*}
M_1&=&\frac{(-1+3n)\sqrt{\pi}\Gamma(-1+2n)}{2\Gamma(\frac{1}{2}+2n)}\nonumber\\
M_2&=& 2^{(-3+4n)}\frac{\Gamma(-1+2n)\Gamma(1+2n)}{\Gamma(4n)}
\end{eqnarray*}
and
\begin{equation*}
	N=\frac{2}{S^2}\sum_{n=1}^{\infty}\frac{S^n(-1)^{n+1}\sqrt{\pi}\Gamma(n)}{n\Gamma(n+{\frac{1}{2}})}
\end{equation*}
\subsection{Evolution equations in parameter space}
From the vanishing conditions of the variational derivatives
$\frac{\delta{\langle{\cal L}\rangle}}{\delta{B}}$,
$\frac{\delta{\langle{\cal L}\rangle}}{\delta{B^*}}$,
$\frac{\delta{\langle{\cal L}\rangle}}{\delta {\tau_0}}$,
$\frac{\delta{\langle{\cal L}\rangle}}{\delta{w}}$,
$\frac{\delta{\langle{\cal L}\rangle}}{\delta{v}}$,
$\frac{\delta{\langle{\cal L}\rangle}}{\delta{q}}$ and $\frac{\delta{\langle{\cal L}\rangle}}{\delta{\sigma}}$ we obtain the following equations
\begin{eqnarray}
&-&4iw\frac{dB^*}{dz}-2iB^*\frac{dw}{dz}+4B^*vw\frac{d\tau_0}{dz}\nonumber-\frac{\pi^2}{6}B^*w\left(w\frac{dq}{dz}\right.\nonumber\\&-&\left.q\frac{dw}{dz}\right)-4B^*w\frac{d\sigma}{dz}+\frac{8}{3}\gamma v\,w|B|^2B^*+\frac{4}{3}\alpha\frac{B^*}{w}\nonumber\\&+&4\alpha B^*v^2w+\frac{\pi^2}{3}\alpha B^*q^2w-\frac{8}{3}\kappa|B|^2B^*w-\frac{4\beta}{G}B^*w\nonumber\\&+&\beta M(2n-1)|B|^{2(2n-2)}B^*w-\frac{4\mu}{S}B^*w\nonumber\\&+&\mu Nn|B|^{2(n-1)}B^*w=0 ,
\label{eq8}
\end{eqnarray}
\begin{eqnarray}
&&\!\!4iw\frac{dB}{dz}+2iB\frac{dw}{dz}+4Bvw\frac{d\tau_0}{dz}-\frac{\pi^2}{6}Bw\left(w\frac{dq}{dz}\right.\nonumber\\&-&\left. q\frac{dw}{dz}\right)-4Bw\frac{d\sigma}{dz}+\frac{8}{3}\gamma vw|B|^2B+\frac{4}{3}\alpha\frac{B}{w}\nonumber\\&+&4\alpha Bv^2w+\frac{\pi^2}{3}\alpha Bq^2w-\frac{8}{3}\kappa|B|^2Bw-\frac{4\beta}{G}Bw\nonumber\\&+&\beta M(2n-1)|B|^{2(2n-2)}Bw-\frac{4\mu}{S}Bw\nonumber\\&+&\mu Nn|B|^{2(n-1)}Bw=0 ,
\label{eq9}
\end{eqnarray}
\begin{equation}
\frac{d}{dz}\left[|B|^2wv\right]=0,
\label{eq10}
\end{equation}
\begin{eqnarray}
&-&\frac{\pi^2}{6}\frac{d}{dz}\left(|B|^2w\right)q-\frac{\pi^2}{2}|B|^2w\frac{dq}{dz}+2i\left(B^*\frac{dB}{dz}\right. \nonumber\\&-&\left. B\frac{dB^*}{dz}\right)+4|B|^2v\frac{d\tau_0}{dz}+\frac{\pi^2}{6}|B|^2q\frac{dw}{dz}-4|B|^2\frac{d\sigma}{dz}\nonumber\\&+&\frac{4}{3}\gamma |B|^4v-\frac{4}{3}\alpha\frac{|B|^2}{w^2}+4\alpha|B|^2v^2+\frac{\pi^2}{3}\alpha|B|^2q^2\nonumber\\&-&\frac{4}{3}\kappa|B|^4-\frac{4\beta}{G}|B|^2+\beta M|B|^{2(2n-1)}-\frac{4\mu}{S}|B|^2\nonumber\\&+&\mu N|B|^{2n}=0,
\label{eq11}
\end{eqnarray}
\begin{equation}
\frac{d\tau_0}{dz}+\frac{1}{3}\gamma|B|^2+2\alpha v=0 ,
\label{eq12}
\end{equation}
\begin{equation}
\frac{dw}{dz}=-2\alpha q,
\label{eq13}
\end{equation}
\begin{equation}
\frac{d}{dz}[w|B|^2]=0 ,
\label{eq14}
\end{equation}
The last equation can also be obtained by multiplying Eq.(\ref{eq8}) by $B$ and Eq.(\ref{eq9}) by $B^*$ and then subtracting the resulting equations.
However, adding Eq.(\ref{eq8})$\times B$ and Eq. (\ref{eq9})$\times B^*$ and then using  Eq.(\ref{eq11})  we get
\begin{eqnarray}
w\frac{dq}{dz}&=&-\frac{4}{\pi^2}\gamma v|B|^2-\frac{8\alpha}{\pi^2w^2}+\frac{4}{\pi^2}\kappa|B|^2\nonumber\\&-&\frac{6 \beta}{\pi^2}(n-1)M|B|^{4(n-1)}\nonumber\\&-&\frac{3 \mu }{\pi^2}(n-1)N|B|^{(2n-2)}.
\label{eq17}
\end{eqnarray}
This is the so-called chirp equation. This equation  in conjunction with  Eq. (\ref{eq12}) leads to
\begin{eqnarray}
\frac{d^2w}{dz^2}&=&\frac{16\alpha^2}{\pi^2w^3}-\frac{8\alpha\kappa}{\pi^2}\frac{|B|^2}{w}+\frac{8\alpha\gamma}{\pi^2}v\frac{|B|^2}{w}\nonumber\\&+&\frac{12\alpha\beta}{\pi^2}M(n-1)\frac{|B|^{4(n-1)}}{w}\nonumber\\&+&\frac{6\alpha\mu}{\pi^2}(n-1)N\frac{|B|^{2(n-1)}}{w}.
\label{eq18}
\end{eqnarray}
Eq. (\ref{eq14}) implies that $w|B|^2$ is a constant (say, $E_0$) which is related to norm ($P$) of the system. The Eq. (\ref{eq10}) immediately demands that $v$ is constant. However, the presence of nonlinear dispersion ($\gamma\ne 0$) in Eq. (\ref{eq12}) indicates that the velocity of the center of motion  can be changed for  any non-zero value of  pulse width due to nonlinear dispersion.
\section{Dynamics of optical soliton}
{In the previous section, we have  derived equations for different parameters and their dependence on the parameters of the system. It is interesting to find the values of system's parameters which can permit stable soliton solution. This can be achieved by analysing the effective potential model \cite{r01,rr0007}. In order of this, we write an effective equation in terms of $w$ from Eq.(\ref{eq18}) as}
\begin{eqnarray}
\frac{d^2w}{dz^2}&=&\frac{16\alpha^2}{\pi^2w^3}-\frac{8\alpha\kappa}{\pi^2}\frac{E_0}{w^2}+\frac{8\alpha\gamma}{\pi^2}v\frac{E_0}{w^2}\nonumber\\&+&\frac{12\alpha\beta}{\pi^2}M(n-1)\frac{E_0^{2(n-1)}}{w^{2n-1}}\nonumber\\&+&\frac{6\alpha\mu}{\pi^2}N(n-1)\frac{E_0^{n-1}}{w^n}.
\label{eq19}
\end{eqnarray}
For the convenience of analysis, we introduce the normalized pulse width $y(z)=\frac{w(z)}{w_0}$ and  write effective potential from Eq. (\ref{eq19}) as
\begin{eqnarray}
\Pi(y)&=&\frac{S_1}{y^2}-\frac{S_2}{y}+\frac{S_3}{y}+\sum\frac{\Omega_n}{y^{2n-2}}+\sum\frac{\Gamma_n}{y^{n-1}}\nonumber\\&-&\left(S_1-S_2+S_3+\sum\Omega_n+\sum \Gamma_n\right),
\label{eq20}
\end{eqnarray}
where 
\begin{eqnarray}
S_1&=&\frac{8\alpha^2}{\pi^2w_0^4}, S_2=\frac{8\alpha\kappa E_0}{\pi^2w_0^3}, S_3=\frac{8\alpha\gamma\, vE_0}{\pi^2 w_0^3}
\label{eq21}
\end{eqnarray}
and
\begin{eqnarray}
\Omega_n=\frac{6\alpha\beta }{\pi^2}M\frac{E_0^{2n-2}}{w_0^{2n}},\,\,\,
\Gamma_n=\frac{6\alpha\mu}{\pi^2}N\frac{E_0^{n-1}}{w_0^{n+1}}.
\label{eq22}
\end{eqnarray}
\begin{figure}[h]
\centering
\includegraphics[width=0.35\textwidth]{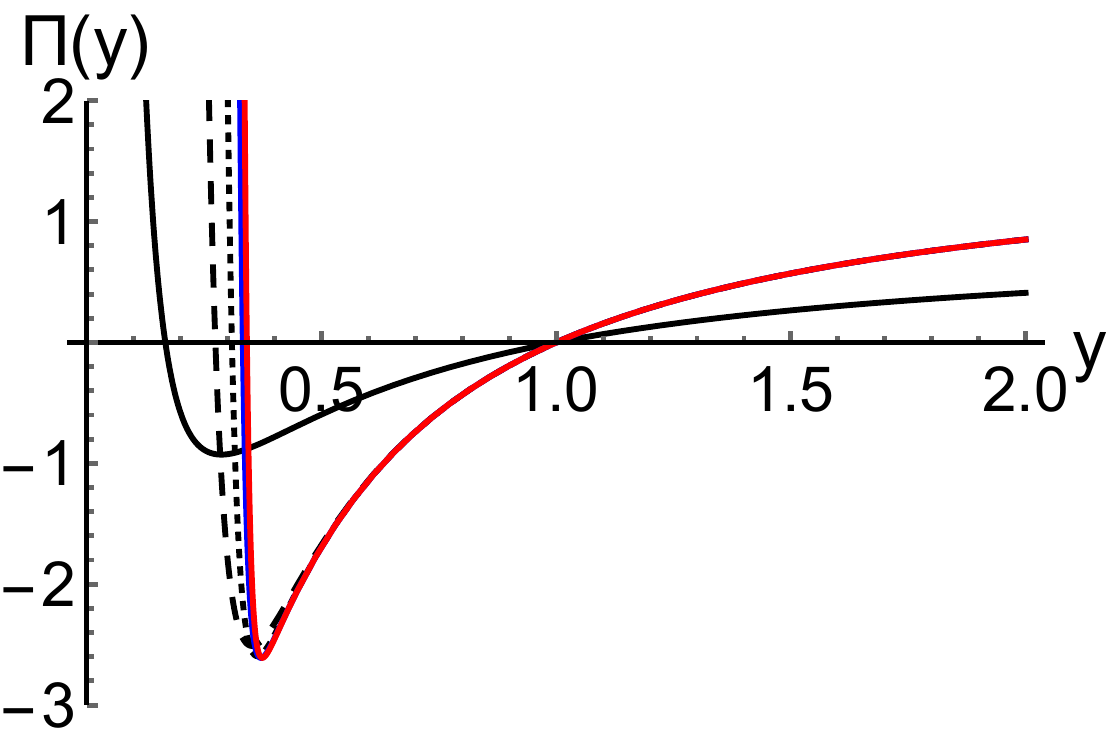}
\includegraphics[width=0.35\textwidth]{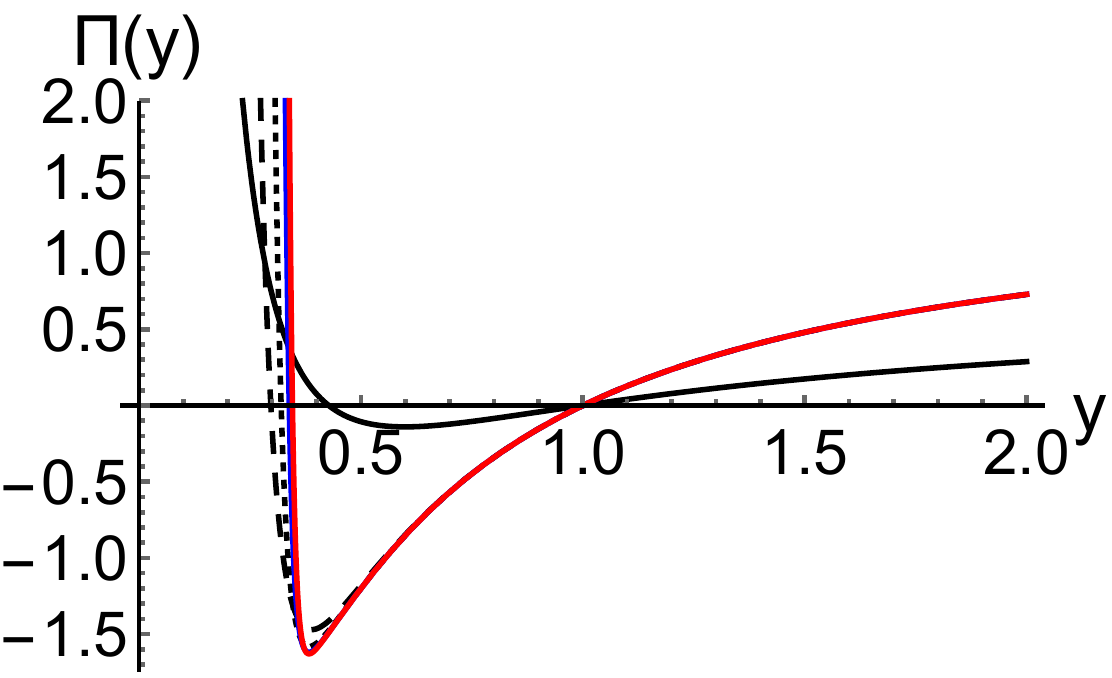}
\caption{Plot of effective potential with pulse width in saturable nonlinear media for $n=1$(black), $n=5$(black dash), $n=9$(black dotted), $n=15$(blue) and  $n=17$(red).  
In all the panels we have taken $\alpha=1.72$, $\beta=3$, $\kappa=4$, $\mu=2$, $w_0=2$ and $G=S=0.27$. Top panel gives the curves for $\gamma=0$ and $v=0$ while we take $\gamma=0.5$ and $v=1.25$ in  bottom panel.}
\label{Fig2}
\end{figure}
\begin{figure}[h]
\centering
\includegraphics[width=0.35\textwidth]{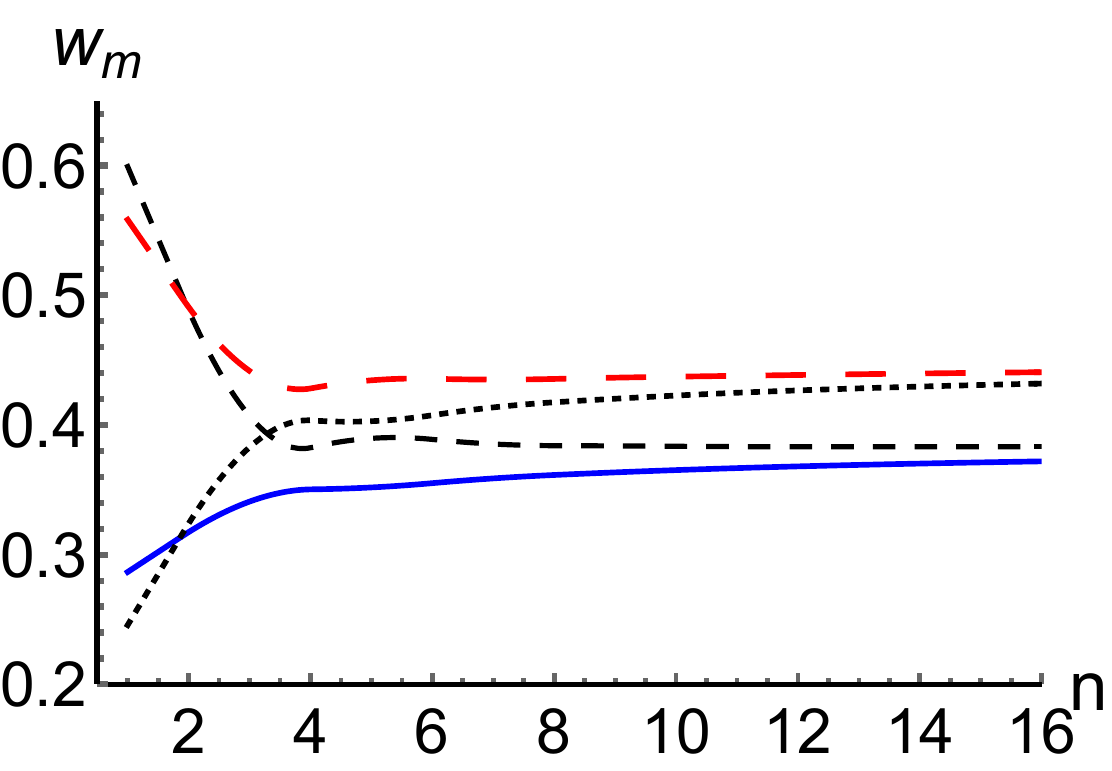}
\hspace*{-0.05cm}
\includegraphics[width=0.35\textwidth]{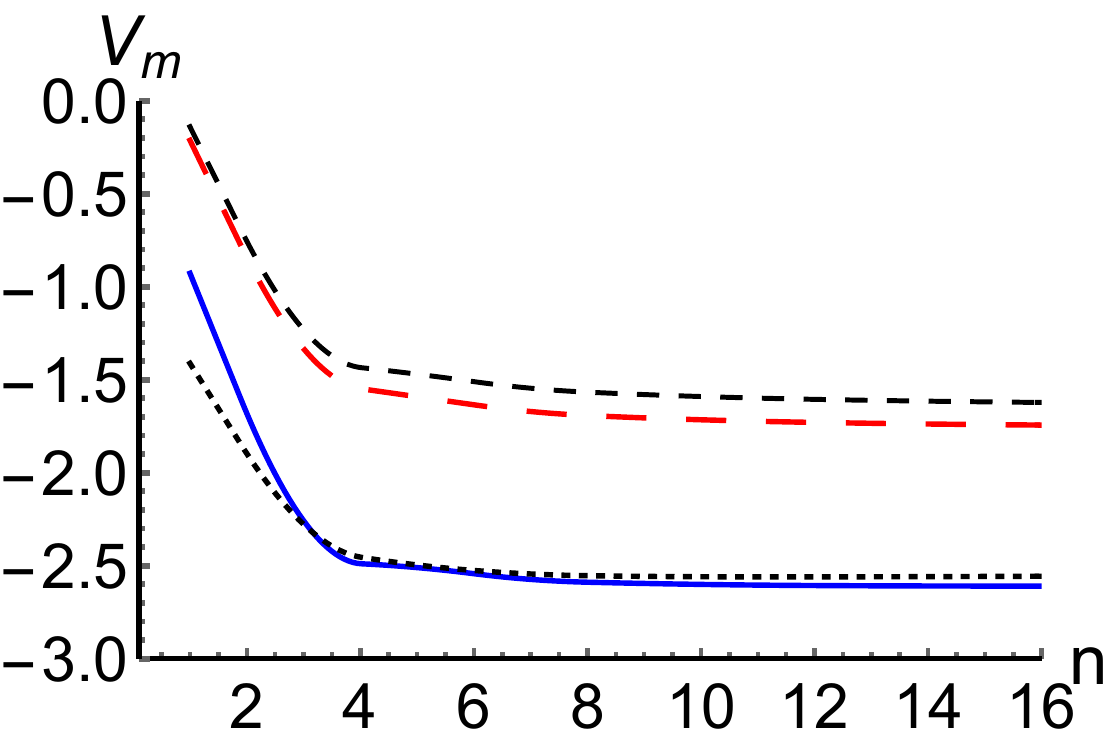}
\caption{Top panel: Effective width corresponding to the potential minimum with the order of quintic saturable nonlinearity.  Bottom panel: Minimum value of effective potential with the order of saturable cubic-quintic  nonlinearity. In all the panels we have taken $\alpha=1.72$, $\beta=3$, $\kappa=4$, $v=1.25$, $w_0=2$ and $G=S=0.27$. Here (i) $E_0=1.5$, $\gamma=0$ (solid blue), (ii)$E_0=1.5$, $\gamma=0.5$ (black dashed), (iii)$E_0=1.75$, $\gamma=0$ (black dotted) and (iv)$E_0=1.75$, $\gamma=1.25$ (red dashed).}
\label{Fig3}
\end{figure}
In Fig. \ref{Fig2}, we plot the variation of effective  potential $\Pi(y)$ with $y$ for different values of $n$ to find the effect of higher order terms arising from saturable nonlinearity. Clearly, the $\Pi(y)$ has a minimum $V_m$ for the chosen values of system's parameters and $n$. The values of width ($w_m$) corresponding to $V_m$ gives stable pulse width.  We know that for $n=1$, the effects of  higher order nonlinearity is  negligible and, therefore, the black solid curve gives effective potential for the fundamental soliton solution. With the increase of $n$ value,  the effect of higher order nonlinearity comes into play. As a result,  energy and width of the soliton start to vary due to interplay among dispersion, Kerr nonlinearity and saturable quintic non-linearity. More specifically, the value of $w_m$ and  $V_m$ decrease with the increase of $n$. In each case, the effect of nonlinear dispersion makes the potential less deeper (bottom panel of Fig. \ref{Fig2}). However, the effect of nonlinear dispersion is dominating in absence of saturable nonlinearities (SNs). Thus the SNs can play an important role in stable pulse propagation.
\begin{figure}[h]
\centering
\includegraphics[width=0.35\textwidth]{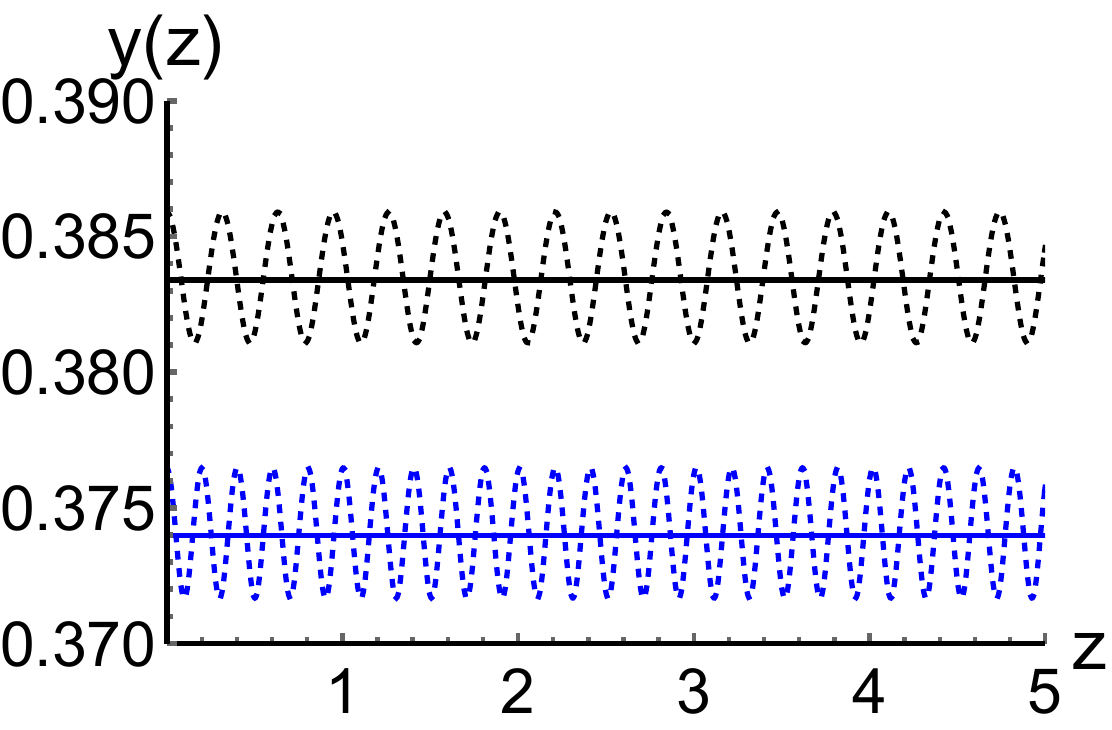}
\caption{variation of effective width as a function of $z$ in cubic-quintic saturable nonlinear media with nonlinear dispersion. Here we have taken $\alpha=1.72$, $\beta=3$, $\kappa=4$ , $G=S=0.27$, $\mu=2$, $E_0=1.5$ and $w_0=2$. (i) $\gamma=0$ and $v=0$ (blue curve), (ii) $\gamma=0.5$ and $v=1.25$ (black curve).}
\label{Fig4}
\end{figure}

To illustrate the insight of saturable limit in present context we  display in Fig. \ref{Fig3} the variation of $w_m$ and $V_m$ with $n$ for (i) $\gamma$=0 and (ii) $\gamma\neq 0$.  The  value of $w_m$ increases with the order $n$ of saturable nonlinearities. However, the growth of  $w_m$ stops at $n=9$. We termed this region as unsaturated region. In this region the dispersive effect dominates over the nonlinear effects resulting a wider pulse. However, in the saturation region ($n \geq 9$) the nonlinear effect can compensate the dispersive effect  and thus the pulse width approaches a constant value. This effect  can be controlled by  $E_0$. More specifically, the value of saturated pulse width or energy for $E_0=1.75$ is higher than that  for  $E_0=1.5$. Therefore, without loss of generality,  we truncate the series arising due to saturable cubic-quintic nonlinearity at $n=19$ and solve $d^2y/dz^2+d\Pi(y)/dy=0$ numerically taking $y(0)=w_m$ and $y'(0)=0$. The variation of normalized pulse width with $z$ is shown Fig.\ref{Fig4}. We see that the pulse width remains unchanged. The value of width for $\gamma\neq 0$ slightly differs from that for $\gamma=0$. However, if the initial pulse width is taken slightly larger than the equilibrium value, pulse width  starts to oscillate in both cases. Their frequencies of oscillations ($\nu$) are not equal. The nonlinear dispersion reduces the value of $\nu$.

\begin{figure}[h]
\centering
\includegraphics[width=0.35\textwidth]{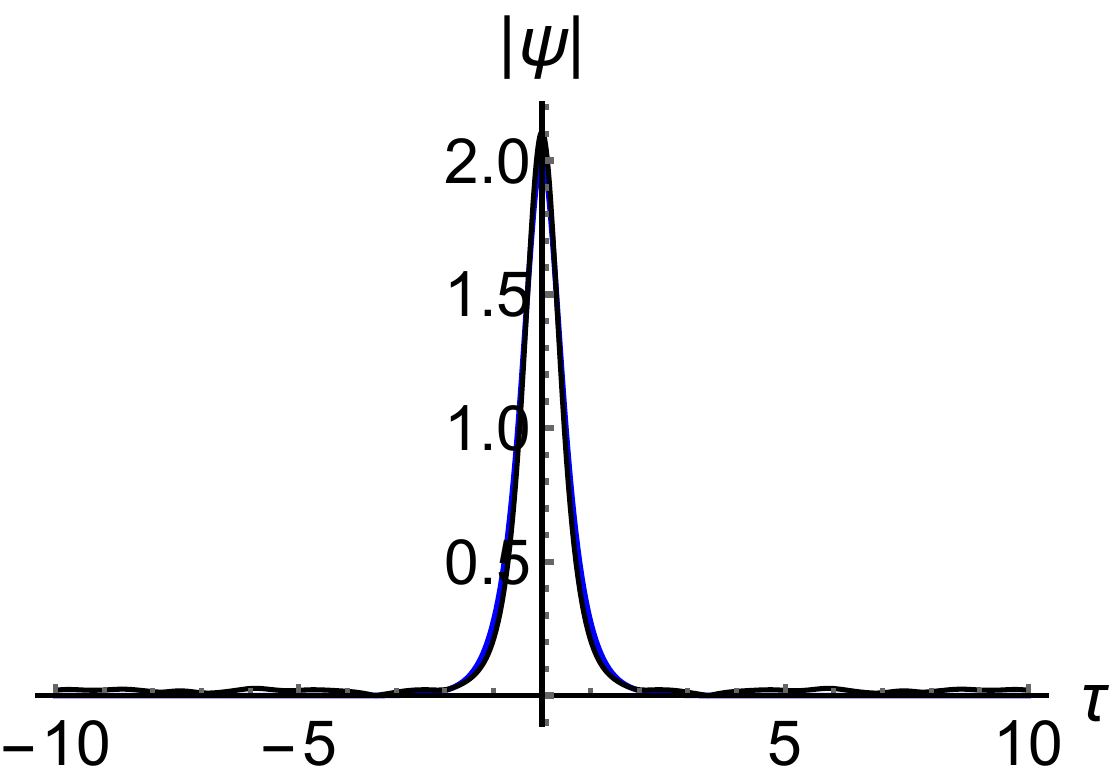}
\includegraphics[width=0.35\textwidth]{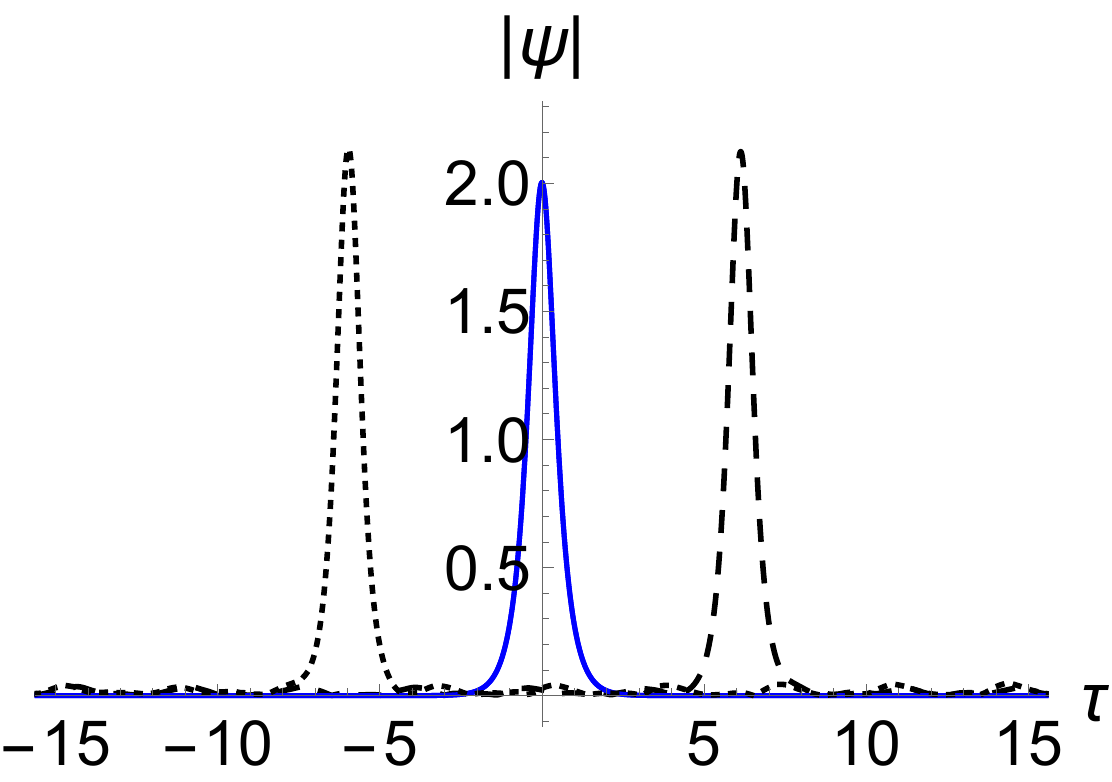}
\caption{Exact numerical solution of Eq.(\ref{eq1}) for (i) $\gamma=0$ (top panel) and (ii) $\gamma=\pm 0.5$ (bottom panel). In top panel, the blue curve represents density profile in calculated at $z=100$ while the black curve gives variational result. In  bottom panel the blue curve gives density profile for $\gamma=0$ while dotted and dashed curve give density profiles for $\gamma=0.5$ (left side) and $\gamma=-0.5$ (right side) calculated at $z=8$ from exact numerical simulation of Eq. (\ref{eq1}). In all the panels we have taken $\alpha=1.72$, $\beta=3$, $\mu=2$, $\kappa=4$ and $G=S=0.27$.}
\label{Fig5}
\end{figure}
We perform exact numerical simulation of Eq.(\ref{eq1}) in the saturated nonlinear regime with initial conditions $\psi(0,\tau)= B(0) \,{\rm sech}[\tau/ w(0)]$ and ${\psi'}(z)|_{z=0}=0$. Here we fixed initial values of $B(0)$ and $w(0)$ at the minimum of the effective potential and calculated density profile at $z=100$ for $\gamma=0$ (top panel, Fig.\ref{Fig5}). We notice that the position and width of the density profile remains unchanged with  $z$ for $\gamma=0$ as predicted by the variational approach. This  indicates that the pulse is dynamically stable. Interestingly, in the presence ($\gamma \neq 0$) of nonlinear dispersion (ND), the soliton starts to move with a constant velocity keeping its shape unchanged (bottom panel of Fig. \ref{Fig5}). The direction of movement depends on the sign of $\gamma$. For a negative value of $\gamma$ it moves towards the right (dashed curve) while it moves towards the left (dotted curve) if $\gamma$ is positive. However, the profile remains  stable even at a larger value of $z$. This result is consistent with the variational prediction in Eq.(\ref{eq12}).

\section{Conclusions}
Pulse  propagation in an optical medium  depends on nonlinear and dispersive responses of the medium. Interestingly, the modern technology has paved the way  of engineering  materials with  manageable  nonlinearity  and  dispersion  in the manufacturing industry by adjusting different parameters/ingredients of the system. In this work we have discussed optical pulse propagation in saturable cubic-quintic nonlinear media with  nonlinear  dispersion. 

We adopted an extended variational method and derived an effective potential for pulse width based on a suitable chosen trial solution. We estimated the trial solution from the shape of the stationary pulse in the zero nonlinear dispersion region. In the variational technique we allow different parameters to vary in order to catch the effect of nonlinear dispersion. Based on a potential model we have examined that energy and width of the soliton attain a limiting value after a certain order of nonlinear terms. However, the saturation limit depends on the power of pulse. We have shown that in the unsaturated regime the cubic-quintic nonlinear medium with nonlinear dispersion support relatively wider solitons. These solitons are stable.  If the soliton is disturbed then its width oscillates  about its equilibrium value. The oscillation frequency is affected by the strength of nonlinear dispersion.  

We  envisaged a direct numerical simulation for the dynamics of soliton both in presence and absence of nonlinear dispersion (ND). In the zero ND regime, the characteristics and position of solitons are found to be unchanged for a long time implying that it is dynamically stable. However, due to  nonlinear dispersion the soliton starts to move with a constant velocity.  The direction of of the movement is determined by sign of the  nonlinear dispersion term.  A similar dynamics can be realized due Raman effect  \cite{rr7} which we will explore somewhere else. 

\subsection*{ACKNOWLEDGEMENT} One of the authors (GS) would like to  acknowledge the funding from the ``Science Research and Engineering Broad, Govt.of India" through Grant No. CRG/2019/000737.

\end{document}